\RequirePackage[l2tabu,orthodox]{nag}

\newif{\ifiopformat}
\iopformatfalse

\newif{\ifhavepgf}
\havepgffalse

\ifiopformat
\documentclass[12pt,a4paper]{iopart}
\else
\documentclass[aps,prb,reprint,showpacs,showkeys,floatfix,a4paper]{revtex4-1}
\fi

\usepackage[utf8]{inputenc}
\usepackage[english]{babel}
\usepackage{csquotes}
\ifiopformat
\usepackage{iopams}
\expandafter\let\csname equation*\endcsname\relax
\expandafter\let\csname endequation*\endcsname\relax
\usepackage[numbers,square,sort&compress]{natbib}
\else
\providecommand{\submitto}{}
\providecommand{\SUST}{}
\usepackage{natbib}
\fi
\usepackage{natmove} %

\usepackage{amssymb}
\usepackage{mathtools}
\usepackage{commath}
\usepackage[mode=math,range-units=single]{siunitx}
\usepackage[caption=false]{subfig}

\ifhavepgf
\usepackage{pgfplots}
\usepackage{tikz}
\usetikzlibrary{plotmarks}
\usetikzlibrary{matrix}
\usetikzlibrary{calc}
\usetikzlibrary{external}
\pgfplotsset{compat=1.5.1}
\pgfplotsset{plot coordinates/math parser=false}
\tikzsetexternalprefix{tikz/}

\tikzset{%
  /tikz/external/only named = true,
}

\tikzset{%
  every pin edge/.append style = {<-,semithick},
  every pin/.append style = {font=\normalsize},
  subplotleft/.style = {baseline,
    trim right={($(current axis.south east) + (0.75em,0)$)},
    trim left={($(current axis.south west) - (3em,0)$)},
    every node/.prefix style = {inner sep=0.3333em}},
  subplotright/.style = {baseline,
    trim right={($(current axis.south east) + (0.75em,0)$)},
    trim left={($(current axis.south west) - (2.5em,0)$)},
    every node/.prefix style = {inner sep=0.3333em}},
  plotcontainer/.style = {inner sep=0, anchor=base east},
}

\pgfkeys{%
  /pgf/number format/set thousands separator={\,},
}
\pgfplotsset{%
  xlabel near ticks,
  ylabel near ticks,
  every axis plot post/.append style = semithick,
  axis line style = thin,
  tick style = thin,
  subplotaxis/.style = {
    tick label style = {font=\footnotesize},
    label style = {font=\small}, %
    tick label style = {inner xsep=0.125em},
    label style = {inner xsep=0.125em},
  },
}

\else

\usepackage{graphicx}
\usepackage{tikzexternal}
\tikzsetexternalprefix{tikz/}

\fi

\newlength{\figureheight}
\newlength{\figurewidth}
\ifiopformat
\else
\fi
\usepackage{ifpdf}
\ifpdf
\usepackage[pdftex]{hyperref} %
\fi
\usepackage[noabbrev]{cleveref} %
\crefname{equation}{\unskip}{\unskip}
\Crefname{equation}{Equation}{Equations}
\usepackage{microtype} %

\usepackage{array}
\usepackage{booktabs}

\ifhavepgf
\usepackage{pgfplotstable}
\pgfplotstableset{%
  include outfiles,
  every table/.append style={outfile={tikz/#1.tex}}, %
  every head row/.style = {before row=\toprule, after row=\midrule},
  every last row/.style = {after row=\bottomrule},
  row style/.style 2 args={
      every row #1 column 0/.style={#2},
      every row #1 column 1/.style={#2},
      every row #1 column 2/.style={#2},
      every row #1 column 3/.style={#2},
      every row #1 column 4/.style={#2},
      every row #1 column 5/.style={#2},
      every row #1 column 6/.style={#2},
      every row #1 column 7/.style={#2},
      every row #1 column 8/.style={#2},
      every row #1 column 9/.style={#2},
      every row #1 column 10/.style={#2},
  }
}
\else
  \let\pgfutilensuremath=\ensuremath
  \let\pgfplotstableresetcolortbloverhangleft=\raggedleft
  \let\pgfplotstableresetcolortbloverhangright=\raggedright
\fi

\newcommand{\mytitle}{Nonequilibrium superconducting thin films with sub-gap and pair-breaking photon illumination}
\newcommand{\tu}[1]{\ensuremath{_{\mathrm{#1}}}}
\newcommand{\unitspace}{\ensuremath{\,\,}}

\ifpdf
\hypersetup{%
  pdfinfo={%
    Title={\mytitle},
    Author={T Guruswamy, D J Goldie and S Withington},
  }
}
\fi

\begin{document}
\title[]{\mytitle}
\ifiopformat
\author{T Guruswamy, D J Goldie and S Withington}
\ead{tg307@mrao.cam.ac.uk}
\address{Quantum Sensors Group, Cavendish Laboratory, University of Cambridge,
J J Thomson Avenue, Cambridge, CB3~0HE, UK}
\else
\author{T Guruswamy}
\email{tg307@mrao.cam.ac.uk}
\author{D J Goldie}
\author{S Withington}
\affiliation{Quantum Sensors Group, Cavendish Laboratory, University of Cambridge,
J J Thomson Avenue, Cambridge, CB3~0HE, UK}
\fi
\date{\today}

\begin{abstract}
We calculate nonequilibrium quasiparticle and phonon distributions for a number of widely-used low transition temperature
 thin-film superconductors
 under constant, uniform illumination by sub-gap probe and pair-breaking signal photons simultaneously.
 From these distributions we calculate
material-characteristic parameters that allow rapid evaluation of an effective quasiparticle temperature
using a simple analytical expression, %
for all materials studied (Mo, Al, Ta, Nb, and NbN) for all photon energies.
We also explore the temperature and energy-dependence of the low-energy
 quasiparticle generation efficiency $\eta$ by pair-breaking signal photons finding $\eta \approx 0.6$ in the limit of thick films at low bath temperatures
that is material-independent. Taking the energy distribution of excess quasiparticles into account,
we find $\eta \to 1$ as the bath temperature approaches the transition temperature in agreement with the
assumption of the two-temperature model of the nonequilibrium response that is appropriate in that regime.
The behaviour of $\eta$ with signal frequency scaled by the superconducting energy gap is also shown
to be material-independent, and is in qualitative agreement with recent experimental results.
An enhancement of $\eta$ in the presence of sub-gap (probe) photons is shown
to be most significant at signal frequencies near the superconducting gap frequency and arises due to multiple photon absorption events that increase the average energy of excess quasiparticles above that in the absence of the probe.
\end{abstract}
\submitto{\SUST}
\maketitle

\section{Introduction}
The energy distribution of the quasiparticle excitations of a superconductor determines its electrical~\cite{Mattis1958}
and thermal transport properties~\cite{Bardeen1959}.
A theoretical understanding of devices including kinetic inductance detectors (KIDs)~\cite{Zmuidzinas2012,Baselmans2012,Vardulakis2008},
superconducting tunnel junctions~\cite{Friedrich2008}, superconducting qubits~\cite{DiCarlo2010,Hofheinz2008,Schoelkopf2008}, SQUID based parametric amplifiers and mixers~\cite{Irwin2004}, and quantum capacitance detectors~\cite{Bueno2010}
requires a description of the quasiparticle system response.
Therefore, a method for calculating the excited, possibly nonequilibrium distribution of quasiparticles is central to any device model.

Nonequilibrium superconducting detectors rely on pair-breaking to create their detected signal.
We are particularly interested in thin-film superconductors fabricated on a dielectric substrate
that is cooled to a sufficiently low temperature (the bath temperature) $T_b$.
Typically $T_b\sim 0.1\,T_c$ where $T_c$ is the superconducting transition temperature.
An interacting signal photon of energy
$h\nu\tu{signal}$, where $h$ is Planck's constant and $\nu\tu{signal}$ the signal frequency, directly breaks a condensate pair in the superconductor provided
$h\nu\tu{signal} \ge 2\Delta$ where $\Delta$ is the superconducting energy gap. Due to the high density of states close to
the gap, and for moderate energy photons, the interaction creates a distribution of excited quasiparticles with peaks\cite{Ivlev1973} at $E=\Delta$
and $E=h\nu\tu{signal} - \Delta$ where $E$ is the quasiparticle energy. $E$ is related to the underlying Bloch state
 energy $\epsilon$ by $E=\sqrt {\epsilon^2+\Delta^2}$.
 Higher energy signal photons will directly release an atomic electron via the photoelectric effect. In this
case effects associated with localized heating and gap-reduction or ``hotspots'' may become important.
In this work we concentrate on modelling the detection of moderate energy photons $h\nu\tu{signal} < \Omega_D$ where these effects are insignificant. The photon
 energies considered are particularly important to understand the responsivity, sensitivity and signal-to-noise of
currently-deployed and planned astronomical instruments performing measurements in the
frequency window of
\SIrange{0.1}{10}{THz}. In the case of KIDs, the nonequilibrium state created by the interaction is monitored using a probe
of energy $h\nu\tu{probe} \ll 2\Delta $, where $\nu\tu{probe}$ is the probe frequency, so an additional drive term needs to be included in the
detailed analysis.

 An important consideration
in any nonequilibrium superconducting detector operating at low reduced temperatures
$T_b/T_c\ll 1$ is that the detected signal is most influenced by the presence of \emph{low-energy}
quasiparticles because the relaxation of the primary excitation occurs on time-scales that are
much shorter than the ultimate relaxation time of the low-energy excess. (Here we define low-energy quasiparticles to have
$E<3\Delta$.)
We assume the signal photons interact with 100\% efficiency i.e. we
 ignore for example any optical coupling efficiencies.
A primary excitation with $E=h\nu\tu{signal} - \Delta$ relaxes towards the gap emitting a phonon. If this
emitted phonon has energy $\Omega \ge 2\Delta$,
an additional pair may be broken enhancing the number of low-energy excess quasiparticles and thus the
detected signal. But in a thin-film this emitted phonon may be lost into the substrate, reducing the number of low-energy excitations.
The average number $m$ of low-energy quasiparticles resulting from the interaction of a single high-energy photon
can be quantified %
in terms of a quasiparticle generation efficiency (or quasiparticle yield) $\eta$.
If the low-energy quasiparticles have average energy $\langle E_{qp}\rangle $ then the
efficiency $\eta=m \langle E_{qp}\rangle/ h\nu\tu{signal} $. Often it is assumed $\langle E_{qp}\rangle =\Delta$~\cite{Zehnder1995}.

Phonon loss means $\eta\le 1$.
Most existing work assumes a value of $\eta \approx 0.6$ for all materials, photon frequencies and bath temperatures~\cite{Zmuidzinas2012}.
At high temperatures $T\approx T_c$ our description of the energy-cascade no longer applies\cite{Semenov_review, Semenov_1995_prb}
 because thermal phonon energies become
comparable with $\Delta$, and scattering and recombination occur on comparable timescales. Then
 a two-temperature model describing the quasiparticle and phonon systems that assumes
 $\eta=1$ is most appropriate.
Here we show that the full nonequilibrium calculation yields $\eta\to 1 $ as $T_b/T_c\to 1$ in agreement with
the assumption of the two-temperature model.

Previous work
modelled the effect of incoming energy at low temperatures with an effective temperature~\cite{Parker1975}
or as an effective chemical potential~\cite{Owen1972} for the quasiparticle distribution.
To go beyond this analysis, \citet{Chang1977,Chang1978} derived a set of
coupled nonlinear kinetic equations describing the interactions of quasiparticles and phonons to find their
respective energy distributions $f(E)$ and $n(\Omega)$.
They solved for the nonequilibrium distributions resulting from various drive terms, including photon and phonon injection.
A number of investigations have explored the effect of very high energy photons (x-ray or optical photons)
on infinite superconductors~\cite{Kurakado1982,Rando1992} and thin films~\cite{Hijmering2009} calculating
 a quasiparticle generation efficiency of $\eta \approx 0.6$, ignoring $\Omega \ge 2\Delta$ phonon loss.
Another approach~\cite{Zehnder1995} considered the time-evolution following
 local energy deposition into the thin film.
\Citet{Kozorezov2000} considered the energy
 downconversion process after absorption of a high energy photon for a
variety of materials and concluded that the materials can be categorized into three different classes,
with the low energy-gap superconductors all having $\eta \approx 0.6$.

We have previously reported~\cite{Goldie2013} a numerical approach that solves the coupled kinetic equations describing the
quasiparticle and phonon distributions $f(E)$ and $n(\Omega)$ for a superconducting thin-film driven by a
sub-gap probe~\cite{Goldie2013}, and including the effect of an additional pair-breaking signal~\cite{Guruswamy2014},
for Al thin-films at low temperatures $T \sim 0.1\,T_c$.
We have also reported
detailed
calculations of KID characteristics (quality factors and quasiparticle lifetimes as functions of readout power and $T_b$)
 that were compared to precise measurements of Al resonator behaviour finding good agreement.~\cite{DeVisser2013a}
More recent measurements on KIDs~\cite{DeVisser2014} of the quasiparticle number and
recombination time-dependence on readout power appear to show that they
must be explained using nonequilibrium quasiparticle distributions.
Other recent measurements with a Ta KID~\cite{Neto2014} report a
detector response that has the same qualitative
features as the energy dependence of $\eta$ for Al described in~\citet{Guruswamy2014} %
Here we apply the method to
a number of other technologically important low temperature superconductors and discuss the physics involved.
We
concentrate in particular on the refractory elements and
 extend the earlier analysis to bath temperatures approaching the superconducting critical temperature $T_c$.
We study Mo, Al, Ta (in its higher $T_c\sim \SI{4.4}{K}$ form),
Nb and NbN. In each instance we use measured material properties
 that characterize films that we deposit by sputtering under ultra-high-vacuum.
Our results are applicable to many different device designs,
as only the phonon escape time to the substrate $\tau_l$ is device-geometry dependent.
The method assumes that phonon pair-breaking, quasiparticle recombination,
and electron-phonon scattering are the only significant interaction processes.
This assumption is investigated in \cref{sec:scattering} for typical thin-films intended for detector applications.
\Cref{sec:Methods}
reviews superconducting parameters for the materials discussed, considers the relative contribution from electron-electron scattering for the different
materials, describes the numerical method, and also describes how $\eta$
is calculated for a pair-breaking signal.
 \Cref{sec:coolingmodel} presents an analytical superconductor cooling model that can be used to calculate the effective
quasiparticle temperature $T_N^*$ for both sub-gap and pair-breaking photons.
\Cref{sec:results} describes detailed numerical results for both sub-gap and pair-breaking photon interactions including
the effect of the effect of changes in the bath temperature. \Cref{sec:conclusions}
summarizes the work done.

\section{Methods}\label{sec:Methods}
\subsection{Superconducting parameters}
The superconducting parameters needed for modelling are the characteristic quasiparticle lifetime
\begin{equation}
  \tau_0 = \frac{Z_1(0) \hbar}{2 \pi b (k_B T_c)^3} ,
\end{equation}
and the characteristic phonon lifetime
\begin{equation}
  \tau_0^\phi = \frac{\hbar N\tu{ion}}{4 \pi^2 N(0)\tu{bs} \langle\alpha^2\rangle \Delta_0} ,
  \label{Eq:tau_}
\end{equation}
as defined by \citet{Kaplan1976} %
$N(0)$ and $N(0)\tu{bs}$ are the mass-enhanced
 and bare single-spin density of states at the Fermi energy $\epsilon_F$ respectively, where
 $N(0)= Z_1(0)N(0)\tu{bs}$, $N\tu{ion}$ is the ion density, and $\Delta_0 $ the zero temperature superconducting energy gap.
$Z_1(0)$ is the electron-phonon renormalization factor $Z_1(0) = 1 + \lambda$,
where $\lambda = 2\int_0^\infty \dif\Omega \, \alpha^2(\Omega) F(\Omega) / \Omega$
 is the dimensionless electron-phonon coupling strength.
 $\alpha^2(\Omega)F(\Omega)$ is the Eliashberg function, $\alpha^2(\Omega)$ is the electron-phonon interaction matrix element and $F(\Omega)$ is the phonon density of states.
The material-dependent parameter $b$ is calculated within a Debye model
 such that $b\Omega^2=\alpha^2(\Omega)F(\Omega)$ for low phonon energies $\Omega \sim 2\Delta_0$.
The averaged value of the interaction matrix element is
 $\langle \alpha^2 \rangle= 1/3 \int_0^\infty \dif\Omega \, \alpha^2(\Omega) F(\Omega)$.
Values of $b$ and $\langle \alpha^2 \rangle$ for Al, Ta and Nb are already tabulated~\cite{Kaplan1976}.
For Mo we use the point-contact measurements of \citet{Caro1981}
 while for NbN we use the tunnelling data of \citet{Kihlstrom1985} and the transport property measurements of \citet{Semenov2009}
We note that the precision of our calculations of $b$ are dependent on the accuracy within which we can interpret the relevant low-frequency
 parts of the reported $\alpha^2(\Omega) F(\Omega)$ data.

\begin{table}[tbh]
  \centering
  \ifhavepgf
  \pgfplotstabletranspose[colnames from=Material, input colnames to=Material]{\materialpropstable}{tables/table1.dat}
  \pgfplotstabletypeset[%
    fixed, precision=2,
    columns = {Material, Mo, Al, Ta, Nb, NbN},
    column type=r,
    columns/Material/.style = {string type, column type=l,
      string replace={Delta0}{$\Delta_0 \unitspace (\si{\micro eV})$},
      string replace={GapFreq}{$\nu\tu{gap} \unitspace (\si{\giga Hz})$},
      string replace={Tc}{$T_c / \si{K}$},
      string replace={b}{$b \unitspace ( 10^{-4}\si{\milli eV^{-2}} )$},
      string replace={a}{$\langle \alpha^2 \rangle \unitspace ( \si{\milli eV} )$},
      string replace={lambda}{$\lambda$},
      string replace={tau0}{$\tau_0 \unitspace ( \si{ns} )$},
      string replace={tau0phi}{$\tau_0^\phi \unitspace ( \si{ps} )$},
      string replace={ratio}{$\tau^\phi\tu{ec}/\tau_0^\phi$},
    },
    row style = {1}{precision=0}
  ]{\materialpropstable}
  \else
  \begin {tabular}{lrrrrr}%
\toprule Material&Mo&Al&Ta&Nb&NbN\\\midrule %
$\Delta _0 \unitspace (\si {\micro eV})$&\pgfutilensuremath {140}&\pgfutilensuremath {180}&\pgfutilensuremath {700}&\pgfutilensuremath {1\,470}&\pgfutilensuremath {2\,560}\\%
$\nu \tu {gap} \unitspace (\si {\giga Hz})$&\pgfutilensuremath {68}&\pgfutilensuremath {87}&\pgfutilensuremath {339}&\pgfutilensuremath {711}&\pgfutilensuremath {1\,240}\\%
$T_c / \si {K}$&\pgfutilensuremath {0.92}&\pgfutilensuremath {1.18}&\pgfutilensuremath {4.6}&\pgfutilensuremath {9.67}&\pgfutilensuremath {16.8}\\%
$b \unitspace ( 10^{-4}\si {\milli eV^{-2}} )$&\pgfutilensuremath {2.28}&\pgfutilensuremath {3.17}&\pgfutilensuremath {17.3}&\pgfutilensuremath {40}&\pgfutilensuremath {47}\\%
$\langle \alpha ^2 \rangle \unitspace ( \si {\milli eV} )$&\pgfutilensuremath {1.62}&\pgfutilensuremath {1.93}&\pgfutilensuremath {1.38}&\pgfutilensuremath {4.6}&\pgfutilensuremath {4.99}\\%
$\lambda $&\pgfutilensuremath {0.42}&\pgfutilensuremath {0.43}&\pgfutilensuremath {0.69}&\pgfutilensuremath {1.84}&\pgfutilensuremath {1.46}\\%
$\tau _0 \unitspace ( \si {ns} )$&\pgfutilensuremath {1\,310}&\pgfutilensuremath {438}&\pgfutilensuremath {1.78}&\pgfutilensuremath {0.15}&\pgfutilensuremath {0.02}\\%
$\tau _0^\phi \unitspace ( \si {ps} )$&\pgfutilensuremath {231}&\pgfutilensuremath {260}&\pgfutilensuremath {22.7}&\pgfutilensuremath {4.17}&\pgfutilensuremath {5.98}\\%
$\tau ^\phi \tu {ec}/\tau _0^\phi $&\pgfutilensuremath {1.04}&\pgfutilensuremath {1.04}&\pgfutilensuremath {0.91}&\pgfutilensuremath {0.94}&\pgfutilensuremath {1.01}\\\bottomrule %
\end {tabular}%

  \fi
  \caption{Characteristic quasiparticle and phonon lifetimes, and associated parameters.
Data from \citet{Gladstone1969}, \citet{Kaplan1976} or \citet{Zehnder1995} unless otherwise specified.
For Mo and NbN we use measurements of $\alpha^2(\Omega)F(\Omega)$ from \citet{Caro1981}
 and \citet{Kihlstrom1985} respectively.
$\tau^\phi\tu{ec}$ is the value of $\tau_0^\phi$ required for energy conservation calculated from \cref{eq:prefactors}.
$\nu\tu{gap} = 2\Delta_0/h$ is the gap frequency.}
  \label{tab:materialparams}
\end{table}

The characteristic times calculated are listed in \cref{tab:materialparams}. %
We assume the weak-coupling relationship $\Delta_0 = 1.76\,k_B T_c$ for all materials to determine $T_c$ in \cref{tab:materialparams}.
The assumption is reasonable for Mo Al and Ta, but gives a slightly higher $T_c$ compared to experiment for Nb (typically $T_c \sim \SI{9.3}{K}$)
and for NbN ($T_c\sim \SI{15}{K})$.
However we do not expect the assumption to significantly affect the physics of the conclusions drawn; the differences introduced by strong coupling are reduced because of the short lifetimes of states with energy $E \gg \Delta$ \cite{Kaplan1976,Chang1977}.
$\tau^\phi\tu{ec}$ is the value of $\tau_0^\phi$ required for energy conservation calculated from \cref{eq:prefactors}
and discussed in more detail in \cref{subsec:methods}.
\subsection{Electron-electron vs electron-phonon scattering} \label{sec:scattering}
Insight into the contribution from electron-electron $e-e$ scattering in
the down-conversion process is most-easily found by considering the relavant scattering rates $T_c$.
For the electron energies of interest $\epsilon \ll \Omega_D$,
 we assume that electron-phonon ($e-p$) inelastic
 scattering dominates over ($e-e$) scattering.
We previously investigated~\cite{Guruswamy2014} to what extent this assumption is valid for low-resistivity Al films.
Now we consider the other low-$T_c$ superconductors using material parameters typical of those that we measure experimentally for the thin-films we deposit by magnetron-sputtering in ultra-high vacuum.

\begin{figure}[tbh]
  \centering
  \tikzsetnextfilename{paper_figure1}
  \input{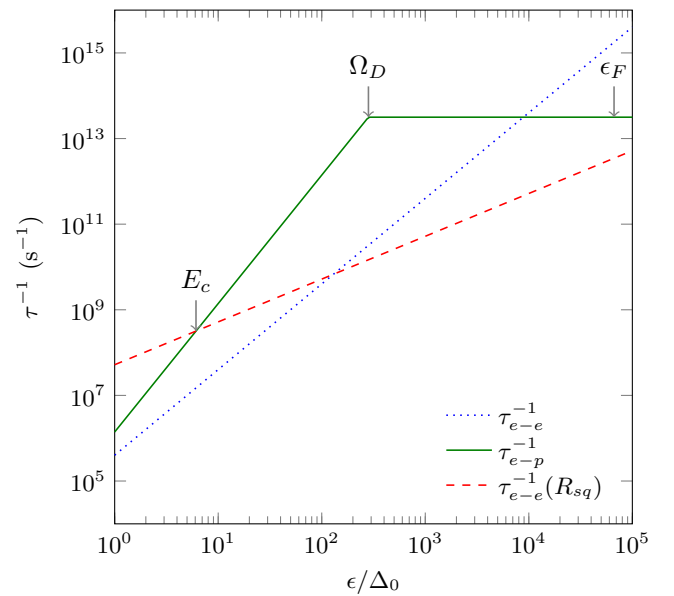}
  \caption{Energy dependence of normal state scattering rates in Mo.
Electron-phonon rate $\tau_{e-p}^{-1}(\epsilon)$ (solid green), clean limit electron-electron
rate $\tau_{e-e}^{-1}(\epsilon)$ (dotted blue), and thin-film resistive $\tau_{e-e}^{-1}(\epsilon,R_{sq})$ (dashed red).
Calculations are for a \SI{40}{nm} Mo film with $\rho = \SI{9.2E-8}{\ohm.m}$.}
  \label{fig:scatteringrates}
\end{figure}

In the normal-state the $e-p$ scattering rate $\tau_{e-p}^{-1}$ is~\cite{Kozorezov2011}
\begin{equation}
  \tau_{e-p}^{-1}(\epsilon) = \frac{2\pi}{\hbar}\int_0^\epsilon \dif\Omega \, \alpha^2(\Omega) F(\Omega) .
\end{equation}
Assuming $\alpha^2(\Omega)$ is constant at low energies, a Debye model for phonons~\cite{Kaplan1976} leads to
\begin{equation}
  \tau_{e-p}^{-1}(\epsilon) = \begin{cases}
    \dfrac{\epsilon^3}{3 \tau_0 (k_B T_c)^3} & \text{if } \epsilon < \Omega_D \vspace{0.5em} \\
    \dfrac{\Omega_D^3}{3 \tau_0 (k_B T_c)^3} & \text{if } \epsilon \ge \Omega_D
  \end{cases} .
\end{equation}
The electron-electron scattering rate in clean films is estimated with the
 Landau-Pomeranchuk formula~\cite{Kozorezov2000}
\begin{equation}
  \tau_{e-e}^{-1}(\epsilon) = \frac{\epsilon^2}{\hbar\epsilon_F}\frac{r_s^{1/2}}{7.96} ,
\end{equation}
where $r_s$ is the radius containing one electron charge divided by the Bohr radius (approximated as $1$). For thin resistive films the $e-e$ rate can be
significantly enhanced so that~\cite{Gershenzon1990}
\begin{equation}
  \tau_{e-e}^{-1}(\epsilon,R_{sq}) = \frac{e^2 R_{sq} \epsilon}{2 \pi^2 \hbar^2} \ln^{-1} \frac{\hbar \pi}{e^2 R_{sq}} ,
  \label{Eq:dirty_ee_rate}
\end{equation}
where $R_{sq} = \rho/d$ is the sheet resistance of the thin-film, $\rho$ is the resistivity and $d$ is the thickness.
\footnote{This differs by a factor of $\pi$ from the expression given in \citet{Altshuler1985}} %
These normal-state calculations significantly overestimate
 the $e-e$ scattering rates in the superconducting state at low temperatures $T_b\ll T_c$~\cite{Sergeev1996}.
\Cref{fig:scatteringrates} plots the normal-state scattering rates for a $40$~nm thick Mo film
and we have used the thin-film resistive $e-e$ scattering rate \cref{Eq:dirty_ee_rate}.
We identify the following characteristic energies for each material: $\epsilon_F$; $\Omega_D$;
and a low-energy crossover $E_c$, below which the $e-e$ scattering rate exceeds the $e-p$ rate.
$E_c/\Delta$ is highest for Mo for the material parameters chosen, as shown in \cref{tab:materialenergies}.

\begin{table}[tbh]
  \centering
  \ifhavepgf
  \begin{filecontents*}{tables/table2.dat}
Material    Delta0    rho d   eF  eD  e2
Al  180 0.8 35  11.6    0.0362  1.42
Ta  700 4.1 100 9.5 0.0207  0.233
Nb  1470    8.8 100 6.18    0.0237  0.138
Mo  140 9.2 40  9.32    0.0396  6.1
NbN 2560    250 100 15.6    0.0284  0.298
\end{filecontents*}
\pgfplotstabletypeset[%
    sort, sort key=Delta0, sort cmp=float <,
    columns = {Material, rho, d, eF, eD, e2},
    column type=r,
    columns/Material/.style = {string type, column type=l,
      string replace={Ta}{Ta {\footnotesize (bcc, epi)}}},
    columns/rho/.style = {column name=$\rho \unitspace\! ( 10^{-8} \si{\ohm.m} )$, dec sep align},
    columns/d/.style = {column name=$d \unitspace\! ( \si{nm} )$, dec sep align},
    columns/eF/.style = {column name=$\epsilon_F \unitspace\! ( \si{eV} )$, dec sep align},
    columns/eD/.style = {multiply by=1E3, column name=$\Omega_D \unitspace\! ( \si{\milli eV} )$, dec sep align},
    columns/e2/.style = {column name=$E_c / \Delta_0$, dec sep align},
    every row 1 column Material/.style = {postproc cell content/.append style = {@cell content/.add={}{\cite{DeVisser2012}}}},
    every row 4 column Material/.style = {postproc cell content/.append style = {@cell content/.add={}{\cite{Toth1971,Schwarz1975}}}}
  ]{tables/table2.dat}
  \else
  \begin {tabular}{lr<{\pgfplotstableresetcolortbloverhangright }@{}l<{\pgfplotstableresetcolortbloverhangleft }r<{\pgfplotstableresetcolortbloverhangright }@{}l<{\pgfplotstableresetcolortbloverhangleft }r<{\pgfplotstableresetcolortbloverhangright }@{}l<{\pgfplotstableresetcolortbloverhangleft }r<{\pgfplotstableresetcolortbloverhangright }@{}l<{\pgfplotstableresetcolortbloverhangleft }r<{\pgfplotstableresetcolortbloverhangright }@{}l<{\pgfplotstableresetcolortbloverhangleft }}%
\toprule Material&\multicolumn {2}{c}{$\rho \unitspace \! ( 10^{-8} \si {\ohm .m} )$}&\multicolumn {2}{c}{$d \unitspace \! ( \si {nm} )$}&\multicolumn {2}{c}{$\epsilon _F \unitspace \! ( \si {eV} )$}&\multicolumn {2}{c}{$\Omega _D \unitspace \! ( \si {\milli eV} )$}&\multicolumn {2}{c}{$E_c / \Delta _0$}\\\midrule %
Mo&$9$&$.2$&$40$&$$&$9$&$.32$&$39$&$.6$&$6$&$.1$\\%
Al\cite {DeVisser2012}&$0$&$.8$&$35$&$$&$11$&$.6$&$36$&$.2$&$1$&$.42$\\%
Ta {\footnotesize (bcc, epi)}&$4$&$.1$&$100$&$$&$9$&$.5$&$20$&$.7$&$0$&$.23$\\%
Nb&$8$&$.8$&$100$&$$&$6$&$.18$&$23$&$.7$&$0$&$.14$\\%
NbN\cite {Toth1971,Schwarz1975}&$250$&$$&$100$&$$&$15$&$.6$&$28$&$.4$&$0$&$.3$\\\bottomrule %
\end {tabular}%

  \fi
  \caption{Characteristic material energies related to scattering. $E_c$ is the energy at which the electron-phonon
scattering rate becomes greater than the thin-film resistive electron-electron scattering rate. Data from \citet{Gladstone1969}, \citet{Kaplan1976} or \citet{Kozorezov2000}
unless otherwise referenced. The material resistivities and thicknesses used are measured values
from thin-films deposited by our group. All data is for polycrystaline films except for Ta where we assume parameters typical of the higher-$T_c$ bcc form.}
  \label{tab:materialenergies}
\end{table}

During the energy down-conversion, the inelastic scattering rate is only relevant above energies $E\ge 3\Delta$ where additional pair breaking is possible.
We find that $E_c$ is below $3\Delta$ for all materials except Mo.
\Cref{fig:scatteringrates} shows that $e-p$ scattering
dominates in that case for $\epsilon \approx 6\Delta$ to $10^4\Delta$ for the modelled Mo film, and $e-e$ scattering may contribute above and below this energy range.
However, the $e-e$ scattering rate is considerably suppressed in the superconducting state at low $T/T_c$.\cite{Sergeev1996}
We conclude that there is negligible contribution from $e-e$ scattering for most of these materials, thicknesses, temperatures
and signal photon energies.
For Mo we note that additional $e-e$ scattering may increase the quasiparticle generation
efficiency over the results presented here by providing an additional pair-breaking mechanism especially near $T_c$.

\subsection{Numerical method}\label{subsec:methods}
We solved
the coupled kinetic equations~\cite{Chang1977} describing the quasiparticle and phonon distributions numerically
to find the steady-state driven distributions, $f(E)$
$n(\Omega)$ for the quasiparticles and phonons respectively, using Newton-Raphson iteration.
Drive terms for pair-breaking and sub-gap photons~\cite{Eliashberg1972} were included as necessary and the numeric pre-factors of these terms scaled
 to match the absorbed powers, $P\tu{probe}$ for sub-gap photons and $P\tu{signal}$
for the pair-breaking signal.
The numerical method was iterated until the solutions conserved energy to within $\xi = \num{E-5}$.
This was quantified as $\xi = \sqrt{\xi_{qp-\phi}^2 + \xi_{\phi-b}^2}$, where $\xi_{qp-\phi}$ is
 the relative difference between the photon power absorbed and power flow from quasiparticles to phonons,
 and $\xi_{\phi-b}$ is the relative difference between the photon power absorbed and the power flow from phonons to substrate (held at a temperature $T_b$).
The numerical method is outlined in more detail in \citeauthor{Goldie2013}~\cite{Goldie2013}.

For each material, the temperature dependent energy gap $\Delta(T)$ was calculated using the BCS gap equation,
\begin{equation}
  \frac{1}{N(0)V\tu{BCS}} = \int_{\Delta(T)}^{\Omega_D} \dif E \frac{1 - 2f(E,T)}{\sqrt{E^2 - \Delta(T)^2}} ,
  \label{eq:gap}
\end{equation}
where $f(E,T) = (\exp{(E/k_B T)}+1)^{-1}$ is the Fermi-Dirac distribution, and $N(0)V\tu{BCS}$ is the dimensionless
electron-phonon coupling constant, calculated from the zero temperature form of \cref{eq:gap} as $N(0)V\tu{BCS} = 1/\sinh^{-1}(\Omega_D/\Delta_0)$.

The numerical method converges on a solution that minimizes $\xi$ provided the material parameters satisfy
\begin{equation}
  \frac{2 \pi N(0) \tau_0^\phi \Delta_0 \Omega_D^3}{9 N\tu{ion} \tau_0 (k_B T_c)^3} = 1 ,
  \label{eq:prefactors}
\end{equation}
a consequence of matching the low energy part of the measured Eliashberg function %
 to a Debye-model approximation %
 that determines $b$.
In our previous work we ensured that $T_c$ satisfied \cref{eq:prefactors}\cite{Goldie2013},
but then the $T_c$ obtained did
 not precisely agree with \cref{eq:gap} (although the difference is small $<1\%)$. %
In this work, since we are also interested in the temperature dependence of the response %
 we calculated $T_c$ from \cref{eq:gap}
 and defined an energy-conserving characteristic phonon lifetime $\tau^\phi\tu{ec}$ to satisfy \cref{eq:prefactors}.
Given the difficulty of interpreting the low-energy
 $\alpha^2(\Omega) F(\Omega)$ data that we and indeed \citet{Kaplan1976} and others have noted
 we consider this approach entirely reasonable.
The ratios $\tau^\phi\tu{ec}/\tau_0^\phi$ %
are
 given in \cref{tab:materialparams}. %
Satisfyingly the ratios are within 10\% of unity for all materials studied.
In our calculation of $b$ for Mo,
 our error would be of this order given the difficulty of
 determining the low-energy $\alpha^2(\Omega)F(\Omega)$ from the available data.
 We note that other estimates of the characteristic times exist (see for example \citet{Parlato_2005}) although
 those estimates do not necessarily satisfy the energy-conserving requirement.

\begin{figure}[tbh]
  \centering

  \ifhavepgf
  \tikzexternaldisable
  \newsavebox\mainplot
  \savebox\mainplot{\input{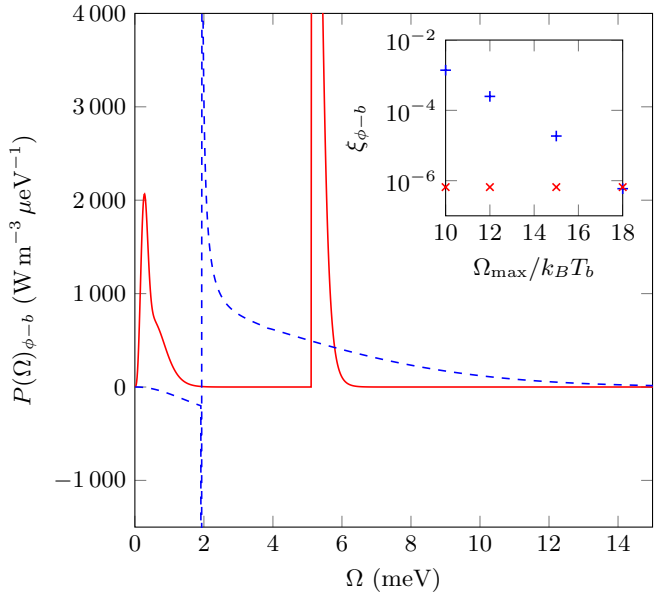}}

  \newsavebox\insetplot
  \savebox\insetplot{%
    \setlength{\figureheight}{0.13\textwidth}%
    \setlength{\figurewidth}{0.13\textwidth}%
%
%
\begin{tikzpicture}

\begin{axis}[%
width=\figurewidth,
height=\figureheight,
scale only axis,
xmin=10,
xmax=18,
xlabel={$\Omega\tu{max} / k_B T_b$},
ymode=log,
ymin=1e-07,
ymax=0.01,
yminorticks=true,
ylabel={$\xi_{\phi-b}$}
]
\addplot [color=blue,only marks,mark=+,mark options={solid},forget plot]
  table[row sep=crcr]{10	0.00138901186498984\\
12	0.000249480925742\\
15	1.86068769037255e-05\\
18	5.90902117243348e-07\\
};
\addplot [color=red,only marks,mark=x,mark options={solid},forget plot]
  table[row sep=crcr]{10	6.55267945713004e-07\\
12	6.55267945713004e-07\\
15	6.55267945713004e-07\\
18	6.55267945713004e-07\\
};
\end{axis}
\end{tikzpicture}%
%
  }
  \tikzexternalenable
  \fi

  \tikzsetnextfilename{paper_figure2}
  \begin{tikzpicture}
  \node[anchor=south west, inner sep=0] (myplot) at (0,0) {\usebox\mainplot};
  \begin{scope}[x={(myplot.south east)}, y={(myplot.north west)}]
    \node[anchor=north east, inner sep=0] at (0.985,0.96) {\usebox\insetplot};
  \end{scope}
  \end{tikzpicture}

  \caption{Phonon-bath power flow $P(\Omega)_{\phi-b}$, at $T_b = 0.1\,T_c$ (solid red)
and $T_b = 0.95\,T_c$ (dashed blue).
Both for NbN, $P\tu{probe} = \SI{2E6}{W.m^{-3}}$, $P\tu{signal} = 0$, $\tau_l/\tau_0^\phi = 1$, and $h\nu\tu{probe} = \SI{16}{\micro eV}$.
Inset of relative phonon-bath power flow difference $\xi_{\phi-b}$ against the energy of
the maximum bin $\Omega\tu{max}/k_B T_b$, at the same bath temperatures $T_b = 0.1\,T_c$ (red $\times$) and $T_b = 0.95\,T_c$ (blue $+$).}
  \label{fig:convergence}
\end{figure}

In the numerical code the energy-bin distribution
 for both the quasiparticles and phonons was also changed when calculating bath temperature dependencies.
As $T_b$ is increased, the phonon-bath power flow $P(\Omega)_{\phi-b}$ above $\Omega = 2\Delta$ increasingly resembles
 the difference between two thermal distributions, and an increasing fraction of power is
 carried by high energy phonons, as shown in \cref{fig:convergence}.
Therefore at high temperatures, the phonon distribution must be accurately calculated to these higher energies.
The energy of the last phonon bin $\Omega\tu{max}$
 significantly affects the energy conservation of the solutions.
\Cref{fig:convergence}~(inset) shows that to have a relative power
 flow difference $\xi_{\phi-b} < \num{E-5}$, required $\Omega\tu{max} > 15\,k_B T_b$ for $T_b/T_c=0.95$.

However, the bin width could not be made too large as the energy gap $\Delta(T)$, readout photon energy $h\nu\tu{p}$,
 and signal photon energy $h\nu\tu{s}$ had to be rounded to the nearest multiple of the bin width, such that
 the photon-induced peaks in the distributions occurred within well-defined energy bins.
We chose to model the phonons with
 $N = 2500$ uniformly-sized bins with maximum phonon energy $\Omega\tu{max}$,
 and similarly the quasiparticles with maximum energy $E\tu{max}=\Omega\tu{max}+\Delta(T)$.
Bin width was therefore variable, depending on material and temperature.
The quasiparticle density of states broadening parameter $\gamma$ was recalculated for each simulation,
 changing for different material and bin width combinations.
It was chosen to minimize the difference between the quasiparticle number calculated as the integral
 $4N(0)\int_\Delta^\infty \dif E \, \rho\tu{BCS}(E,\Delta) f(E,T_b)$, where $\rho\tu{BCS}(E,\Delta) = E/\sqrt{E^2 - \Delta^2}$,
 and the equivalent sum over the discretized distribution with the broadened density of states calculated
 with $E\to E + i \gamma$.
In all cases the approach worked well.

\subsection{Low-energy quasiparticle generation efficiency for a pair-breaking signal} \label{subsec:qpgenefficiency}
For a single pair-breaking photon we have defined a low-energy generation efficiency
$\eta=m \langle E_{qp}\rangle/ h\nu\tu{signal}$.
Here we describe the calculation of $\eta$ for a constant input signal power $P\tu{signal}$
 in a thin-film that may also be driven by a non-direct pair-breaking constant readout power $P\tu{probe}$ (such as used for KID readout).
The absorbed power creates a number of excess quasiparticles $N\tu{excess} = N\tu{signal} - N\tu{probe}$ such that the
 quasiparticle system total energy has increased by $E\tu{excess} = E\tu{signal} - E\tu{probe}$, where $N\tu{signal}$ and
 $E\tu{signal}$ are the quantities for distributions driven by the signal and probe together.
In the steady state, photons absorbed at a rate $\Gamma_\Phi$ create low-energy
 quasiparticles at a
rate $\Gamma_{qp} = m\Gamma_\Phi$.

Solving a modified set of Rothwarf-Taylor rate equations~\cite{Rothwarf1967} for $\Gamma_\Phi$ and $\Gamma_{qp}$ results in~\cite{Guruswamy2014}
\begin{equation}
\eta = \langle{E_{qp}}\rangle \frac{(N\tu{signal}^2 - N\tu{probe}^2)}{P\tu{signal}} \frac{2R}{1+\beta \tau_l} \; ,
\end{equation}
where $R$ and $\beta$ are the distribution-averaged recombination rate and %
pair-breaking rates respectively. %
We define $\langle{E_{qp}}\rangle$
the average energy of the excess low-energy quasiparticles where
\begin{equation}
\langle{E_{qp}}\rangle = \frac{\int_0^\infty E\rho(E)(f(E)-f(E,T))\dif E}{\int_0^\infty \rho(E)(f(E)-f(E,T))\dif E }.
\end{equation}
At low temperatures $\langle{E_{qp}}\rangle \approx \Delta$ as usually assumed: taking account of the detailed energy distributions allows
$\eta$ to be determined at arbitrary temperatures.

\section{Superconductor cooling model} \label{sec:coolingmodel}
We used the steady-state driven distributions that were calculated to determine
parameters for the thin-film superconductor cooling model outlined in \citeauthor{Goldie2013}~\cite{Goldie2013}.
This model is an analytic expression relating
 $T_N^*$ to the
 power flow $P$ between the
quasiparticle and phonon systems (the latter assumed to be the substrate phonons) for a given $T_b \ll T_c$ and material.
\begin{multline}
  P = \frac{1}{\eta(P,\nu)} \;
    \frac{\Sigma_s}{1+\tau_l/\tau_{pb}} \times \\
  \left[
    T_N^* \exp\left(\frac{-2\Delta(T_N^*)}{k_B T_N^*}\right) -
    T_b \exp\left(\frac{-2\Delta(T_b)}{k_B T_b}\right)
  \right]
  . \label{eq:p_model}
\end{multline}
$\Sigma_s$ is a material-dependent constant, $\tau_{pb}$ is the phonon pair breaking time, $\sim \tau_0^\phi$
for $T_b\ll T_c$.
$T_N^*$ is the effective temperature calculated to characterize the driven
 nonequilibrium quasiparticle distributions so that $N\tu{qp} = 4N(0) \int_\Delta^\infty \dif E \, \rho(E) f(E,T_N^*)$.
$\eta(P,\nu)$ depends on the drive (probe or pair-breaking signal). \Citet{Goldie2013}
 showed for a sub-gap probe
 $\eta(P,\nu\tu{probe})\equiv \eta_{2\Delta}$ the fraction of phonon-bath power flow carried by excess phonons
with energy $\Omega > 2\Delta(T_b)$ that depends on $P$, $\nu\tu{probe}$ and $\tau_l$. Here we show that \cref{eq:p_model} can also be
applied for direct
 pair-breaking when $\eta(P,\nu\tu{signal})\equiv \eta$ as defined in \cref{subsec:qpgenefficiency}.

\section{Results} \label{sec:results}
\subsection{Effects of a sub-gap probe}\label{sec:eta_sub}
\begin{figure}[tbh]
  \setlength{\figureheight}{0.34\columnwidth}
  \setlength{\figurewidth}{0.34\columnwidth}

  \centering
  \tikzsetnextfilename{paper_figure3}
  \begin{tikzpicture}
    \matrix [matrix of nodes, ampersand replacement=\&] {%
      |[plotcontainer]| 
%
%
\begin{tikzpicture}[%
subplotleft
]

\begin{axis}[%
width=\figurewidth,
height=\figureheight,
scale only axis,
xmode=log,
xmin=0.01,
xmax=100000000,
xminorticks=true,
xlabel={$P\tu{probe}\unitspace (\si{W.m^{-3}})$},
ymin=0.9,
ymax=1.8,
ylabel={$T_N^*\unitspace (\si{K})$},
subplotaxis, extra description/.code = {\node at (0.3,0.8) {(a)};}
]
\addplot [color=blue,only marks,mark=+,mark options={solid},forget plot]
  table[row sep=crcr]{0.0632455532033676	0.968869317325398\\
0.2	0.972457216677858\\
0.632455532033676	0.981516976105892\\
2	0.99937300975607\\
6.32455532033676	1.02597726869865\\
20	1.05872229153141\\
63.2455532033676	1.09555078170888\\
200	1.1356663697216\\
632.455532033676	1.17880940920571\\
2000	1.22502884827919\\
6324.55532033676	1.27441471528099\\
20000	1.32701989227757\\
63245.5532033676	1.38293524771484\\
200000	1.44263571148608\\
632455.532033676	1.5066511950979\\
2000000	1.57554226783617\\
6324555.32033676	1.64985975453876\\
20000000	1.73025662508791\\
};
\addplot [color=red,solid,forget plot]
  table[row sep=crcr]{0.0648296994142235	0.968869317325398\\
0.204924739273728	0.972457216677858\\
0.647637161991614	0.981516976105892\\
2.04557375433686	0.99937300975607\\
6.4530423721345	1.02597726869865\\
20.3751222212103	1.05872229153141\\
64.34522877095	1.09555078170888\\
203.343417996301	1.1356663697216\\
641.768338141617	1.17880940920571\\
2024.02298552414	1.22502884827919\\
6379.75227859059	1.27441471528099\\
20079.1737286521	1.32701989227757\\
63028.7366285566	1.38293524771484\\
197375.708595732	1.44263571148608\\
616269.087584016	1.5066511950979\\
1917620.40637282	1.57554226783617\\
5943901.43980349	1.64985975453876\\
18346778.0299575	1.73025662508791\\
};
\addplot [color=red,dotted,forget plot]
  table[row sep=crcr]{0.064314133428155	0.968869317325398\\
0.203420113260481	0.972457216677858\\
0.643061708356459	0.981516976105892\\
2.03156159836265	0.99937300975607\\
6.40699686072722	1.02597726869865\\
20.2052796778736	1.05872229153141\\
63.64005424246	1.09555078170888\\
200.113585395165	1.1356663697216\\
625.619840973772	1.17880940920571\\
1939.44775438905	1.22502884827919\\
6200.9789447403	1.27441471528099\\
20027.6894007548	1.32701989227757\\
63682.996522744	1.38293524771484\\
200211.578430522	1.44263571148608\\
624702.513196205	1.5066511950979\\
1940388.35392127	1.57554226783617\\
6011848.33720765	1.64985975453876\\
18634218.0655167	1.73025662508791\\
};
\end{axis}
\end{tikzpicture}%
%
      \&%
      |[plotcontainer]| 
%
%
\begin{tikzpicture}[%
subplotright
]

\begin{axis}[%
width=\figurewidth,
height=\figureheight,
scale only axis,
xmode=log,
xmin=0.01,
xmax=100000000,
xminorticks=true,
xlabel={$P\tu{probe}\unitspace (\si{W.m^{-3}})$},
ymin=0.4,
ymax=1,
ylabel={$\eta_{2\Delta}$},
subplotaxis, extra description/.code = {\node at (0.3,0.2) {(b)};}
]
\addplot [color=blue,only marks,mark=+,mark options={solid},forget plot]
  table[row sep=crcr]{0.0632455532033676	0.975005628495856\\
0.2	0.975443924924252\\
0.632455532033676	0.975323430131888\\
2	0.974871245396832\\
6.32455532033676	0.973938931238511\\
20	0.972350255579964\\
63.2455532033676	0.969518593531971\\
200	0.964429352979118\\
632.455532033676	0.954948312008363\\
3017.01715497075	0.938036717887639\\
6324.55532033676	0.910304767406686\\
20000	0.869131249848212\\
63245.5532033676	0.814363116806586\\
200000	0.751251896221086\\
632455.532033676	0.684555914175185\\
2000000	0.617406243038157\\
6324555.32033676	0.551423021341327\\
20000000	0.487972448172197\\
};
\addplot [color=red,solid,forget plot]
  table[row sep=crcr]{0.0632455532033676	0.967386639217439\\
0.2	0.967386639217439\\
0.632455532033676	0.967386639217439\\
2	0.967386639217439\\
6.32455532033676	0.967386639217439\\
20	0.967386639217439\\
63.2455532033676	0.967386639217439\\
200	0.967386639217439\\
632.455532033676	0.967386639217439\\
3017.01715497075	0.967386639217439\\
};
\addplot [color=red,solid,forget plot]
  table[row sep=crcr]{3017.01715497075	0.967386639217439\\
6324.55532033676	0.927569164648139\\
20000	0.865635426784722\\
63245.5532033676	0.803701688921305\\
200000	0.741767951057888\\
632455.532033676	0.679834213194471\\
2000000	0.617900475331054\\
6324555.32033676	0.555966737467637\\
20000000	0.49403299960422\\
};
\addplot [color=black,dotted,forget plot]
  table[row sep=crcr]{3017.01715497075	0.4\\
3017.01715497075	1\\
};
\end{axis}
\end{tikzpicture}%
%
      \\
      |[plotcontainer]| 
%
%
\begin{tikzpicture}[%
subplotleft
]

\begin{axis}[%
width=\figurewidth,
height=\figureheight,
scale only axis,
xmode=log,
xmin=0.01,
xmax=100000000,
xminorticks=true,
xlabel={$P\tu{probe}\unitspace (\si{W.m^{-3}})$},
ymin=3.51553414951283,
ymax=4.29676396051568,
ylabel={$\tau_{pb}\unitspace (\si{ps})$},
subplotaxis, extra description/.code = {\node at (0.3,0.2) {(c)};}
]
\addplot [color=blue,only marks,mark=+,mark options={solid},forget plot]
  table[row sep=crcr]{0.0632455532033676	4.01079667877648\\
0.2	4.00945119353812\\
0.632455532033676	4.00617850988099\\
2	4.00066532582735\\
6.32455532033676	3.99525070380874\\
20	3.9918355573872\\
63.2455532033676	3.98967803689287\\
200	3.98748801710441\\
632.455532033676	3.98421842127039\\
3017.01715497075	3.9788415408177\\
6324.55532033676	3.97032578815909\\
20000	3.95799556041201\\
63245.5532033676	3.94194248460798\\
200000	3.92304310182069\\
632455.532033676	3.9017504220322\\
2000000	3.87802795258848\\
6324555.32033676	3.85157887513837\\
20000000	3.82203752664604\\
};
\addplot [color=black,dotted,forget plot]
  table[row sep=crcr]{0.01	3.90614905501425\\
100000000	3.90614905501425\\
};
\end{axis}
\end{tikzpicture}%
%
      \&%
      |[plotcontainer]| 
%
%
\begin{tikzpicture}[%
subplotright
]

\begin{axis}[%
width=\figurewidth,
height=\figureheight,
scale only axis,
xmin=0,
xmax=25,
xlabel={$\tau_l/\tau_0^\phi$},
ymin=0.855900850739103,
ymax=1,
ylabel={$\eta_{2\Delta}$},
subplotaxis, extra description/.code = {\node at (0.8,0.2) {(d)};}
]
\addplot [color=blue,only marks,mark=+,mark options={solid},forget plot]
  table[row sep=crcr]{0.5	0.932513779169029\\
1	0.938036717887639\\
2	0.944840933119958\\
5	0.953873867735226\\
10	0.959736970952371\\
15	0.962550566916959\\
25	0.965453781078512\\
};
\end{axis}
\end{tikzpicture}%
%
      \\
    };
  \end{tikzpicture}
  {
  \subfloat{\label{fig:tnvsp}}
  \subfloat{\label{fig:eta2DvsP}}
  \subfloat{\label{fig:taupbvsP}}
  \subfloat{\label{fig:eta2Dvstaul}}
  }
  \caption{\protect\subref*{fig:tnvsp} Effective quasiparticle temperatures $T_N^*$ as a function of $P\tu{probe}$;
full nonequilibrium calculation (blue $+$), and a fit using \protect\cref{eq:p_model}
 (red solid line).
\protect\subref*{fig:eta2DvsP} Fraction of phonon-bath power flow carried by $\Omega > 2\Delta$ excess phonons %
as a function of $P\tu{probe}$; full nonequilibrium calculation (blue $+$), and piecewise fit (red line) to \protect\cref{eq:eta2Dfit}. The dotted vertical line marks $P_0$.
\protect\subref*{fig:taupbvsP} Distribution-averaged phonon pair-breaking time $\tau_{pb}$. The characteristic phonon lifetime $\tau_0^\phi$ is marked with the dotted line.
\protect\subref*{fig:eta2Dvstaul} $\eta_{2\Delta}$ as a function of phonon escape time ratio $\tau_l/\tau_0^\phi$.
The calculations are for Nb, with $P\tu{probe} = \SI{2E3}{W.m^{-3}}$, $h\nu\tu{probe} = \SI{16}{\micro eV}$, and $\tau_l/\tau_0^\phi = 1$.}
  \label{fig:othersvsp}
\end{figure}
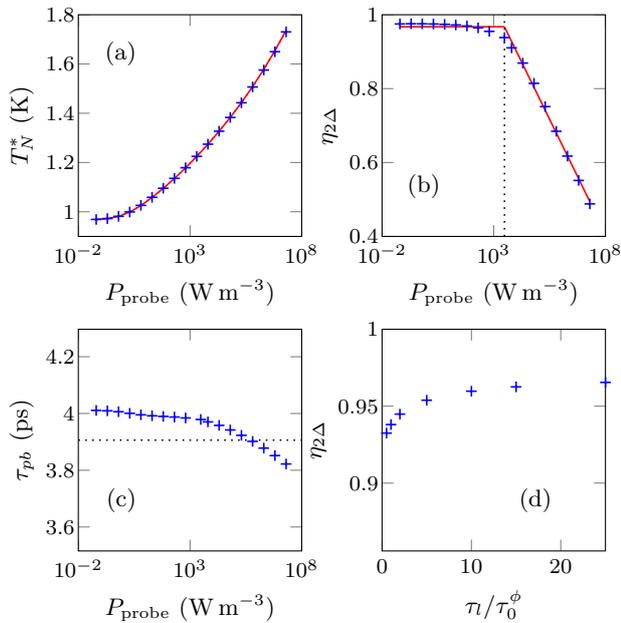

Here we show and discuss results for modelling of the effect of a sub-gap probe. The data-points (crosses) in \cref{fig:tnvsp}
 show
the calculated %
$T_N^*$ from the full non-equilibrium solution as a function of $P\tu{probe}$. The solid line
shows $T_N^*$ evaluated with \cref{eq:p_model} and from this we also calculate
$\Sigma_s$. The calculations are for Nb, with $P\tu{probe} = \SI{2E3}{W.m^{-3}}$, $h\nu\tu{probe} = \SI{16}{\micro eV}$, and $\tau_l/\tau_0^\phi = 1$.
The analytical model is an
 excellent approximation to the full nonequilibrium calculation provided the power-dependence of $\eta_{2\Delta}$ shown in
\cref{fig:eta2DvsP} is taken into account.
The solid line in \cref{fig:eta2DvsP} is a piecewise fit such that
\begin{equation}
  \eta_{2\Delta}(P) = \begin{cases}
                              \eta_0 & \text{if } P \le P_0 \\
                              \eta_1 \ln(P / P_0) + \eta_0 & \text{if } P > P_0.
                            \end{cases} \label{eq:eta2Dfit}
\end{equation}
 $P_0$ characterizes a ``knee'' probe power below which $\eta_{2\Delta}$ is constant and %
 $\eta_1$ characterizes the energy dependence of $\eta_{2\Delta}$ at higher probe powers. Both are material dependent.
\Cref{fig:taupbvsP} shows that $\tau_{pb} \sim \tau_0^\phi$ is a good approximation over the range of interest.
 \Cref{fig:eta2Dvstaul}
 shows that $\eta_{2\Delta}$ has no significant dependence on the phonon escape time $\tau_l$.

\begin{figure}[tbh]
  \setlength{\figureheight}{0.32\columnwidth}
  \setlength{\figurewidth}{0.32\columnwidth}

  \centering
  \tikzsetnextfilename{paper_figure4}
  \begin{tikzpicture}
    \node[anchor=south east, xshift=-1mm] {
%
%
\begin{tikzpicture}[%
subplotleft, trim left={($(current axis.south west) - (4em,0)$)}
]

\begin{axis}[%
width=\figurewidth,
height=\figureheight,
scale only axis,
xmin=0,
xmax=0.5,
xlabel={$h\nu_p / \Delta_0$},
ymode=log,
ymin=1e-15,
ymax=1e-09,
yminorticks=true,
ylabel={$k_B P_0 / (\Sigma_s \Delta_0)$},
subplotaxis, extra description/.append code = {\node at (0.8,0.2) {(a)};}
]
\addplot [color=red,only marks,mark=o,mark options={solid},forget plot]
  table[row sep=crcr]{0.0444444444444445	9.15220088805553e-15\\
0.0888888888888889	4.19462544083348e-14\\
0.133333333333333	2.57888738846427e-13\\
0.177777777777778	2.21035262348926e-12\\
0.266666666666667	5.00089974489146e-11\\
0.355555555555556	1.97598155094639e-10\\
0.526370562259151	1.0999999e-09\\
};
\addplot [color=black!50!green,only marks,mark=square,mark options={solid},forget plot]
  table[row sep=crcr]{0.0114285714285714	5.36196175074591e-14\\
0.0228571428571429	1.36859534213421e-14\\
0.0342857142857143	1.00774603141052e-14\\
0.0457142857142857	1.01465994503378e-14\\
0.0685714285714286	1.49861323833944e-14\\
0.0914285714285714	5.42697386733432e-14\\
};
\addplot [color=blue,only marks,mark=x,mark options={solid},forget plot]
  table[row sep=crcr]{0.0054421768707483	9.75541488974629e-14\\
0.0108843537414966	4.62295385032438e-14\\
0.0163265306122449	1.89644803404255e-14\\
0.0217687074829932	1.51018756797154e-14\\
0.0326530612244898	1.09771034038436e-14\\
0.0435374149659864	1.10309115018508e-14\\
};
\end{axis}
\end{tikzpicture}%
};
    \node[anchor=south west, xshift=-4.5mm] {
%
%
\begin{tikzpicture}[%
subplotright
]

\begin{axis}[%
width=\figurewidth,
height=\figureheight,
scale only axis,
xmin=0,
xmax=0.5,
xlabel={$h\nu_p / \Delta_0$},
ymin=0.2,
ymax=1,
ylabel={$\eta_0$},
subplotaxis, extra description/.append code = {\node at (0.75,0.75) {(b)};}
]
\addplot [color=red,only marks,mark=o,mark options={solid},forget plot]
  table[row sep=crcr]{0.0444444444444445	0.779287690710866\\
0.0888888888888889	0.612596221852822\\
0.133333333333333	0.505687046341003\\
0.177777777777778	0.42894875708492\\
0.266666666666667	0.332861405114083\\
0.355555555555556	0.270537525407793\\
0.533333333333333	0.195363292909432\\
};
\addplot [color=black!50!green,only marks,mark=square,mark options={solid},forget plot]
  table[row sep=crcr]{0.0114285714285714	0.964711082766041\\
0.0228571428571429	0.879352091330929\\
0.0342857142857143	0.828498144679307\\
0.0457142857142857	0.780370894923981\\
0.0685714285714286	0.680129416406068\\
0.0914285714285714	0.612802351008531\\
0.128191007861133	0.12\\
};
\addplot [color=blue,only marks,mark=x,mark options={solid},forget plot]
  table[row sep=crcr]{0.0054421768707483	0.997470046363052\\
0.0108843537414966	0.967386639217439\\
0.0163265306122449	0.940955018793998\\
0.0217687074829932	0.910934783025392\\
0.0326530612244898	0.828652692196853\\
0.0435374149659864	0.780506606644026\\
0.0619592642984879	0.12\\
};
\end{axis}
\end{tikzpicture}%
};
    \node[anchor=north, yshift=1mm, xshift=-5mm] {
%
%
\begin{tikzpicture}[%
subplotleft
]

\begin{axis}[%
width=\figurewidth,
height=\figureheight,
scale only axis,
xmin=0,
xmax=0.5,
xlabel={$h\nu_p / \Delta_0$},
ymin=-0.07,
ymax=-0.02,
ylabel={$\eta_1$},
subplotaxis, scaled ticks=false, tick label style={/pgf/number format/fixed}, extra description/.append code = {\node at (0.8,0.25) {(c)};}
]
\addplot [color=red,only marks,mark=o,mark options={solid},forget plot]
  table[row sep=crcr]{0.0444444444444445	-0.0359862269585353\\
0.0888888888888889	-0.0299236118779654\\
0.133333333333333	-0.0266178685975955\\
0.177777777777778	-0.0257696723842347\\
0.266666666666667	-0.0253151346366859\\
0.355555555555556	-0.02168695898928\\
0.533333333333333	-0.0154214365117164\\
};
\addplot [color=black!50!green,only marks,mark=square,mark options={solid},forget plot]
  table[row sep=crcr]{0.0114285714285714	-0.054528319492585\\
0.0228571428571429	-0.0431310167749742\\
0.0342857142857143	-0.0395308265642134\\
0.0457142857142857	-0.0368879135332304\\
0.0685714285714286	-0.0316477339952407\\
0.0914285714285714	-0.0313014801097806\\
0.115236088411495	-0.015\\
};
\addplot [color=blue,only marks,mark=x,mark options={solid},forget plot]
  table[row sep=crcr]{0.0054421768707483	-0.0604227719970024\\
0.0108843537414966	-0.0537949611954491\\
0.0163265306122449	-0.0475510430625073\\
0.0217687074829932	-0.0453140924936018\\
0.0326530612244898	-0.040011903326224\\
0.0435374149659864	-0.0373204399244385\\
0.0565567450861721	-0.015\\
};
\end{axis}
\end{tikzpicture}%
};
  \end{tikzpicture}
  {
    \subfloat{\label{fig:P0vshnup}}
    \subfloat{\label{fig:eta0vshnup}}
    \subfloat{\label{fig:eta1vshnup}}
  }
  \caption{Parameters for piecewise fits to $\eta_{2\Delta}$ for Nb (blue $\times$), Ta (green $\square$), and Al (red $\bigcirc$) calculated at a range of frequencies $h\nu\tu{probe}$.
\protect\subref*{fig:P0vshnup} $P_0$ against scaled frequency.
\protect\subref*{fig:eta0vshnup} $\eta_0$ (constant section) against scaled frequency.
\protect\subref*{fig:eta1vshnup} $\eta_1$ (slope for linear section) against scaled frequency.}
  \label{fig:fitparamsvshnup}
\end{figure}
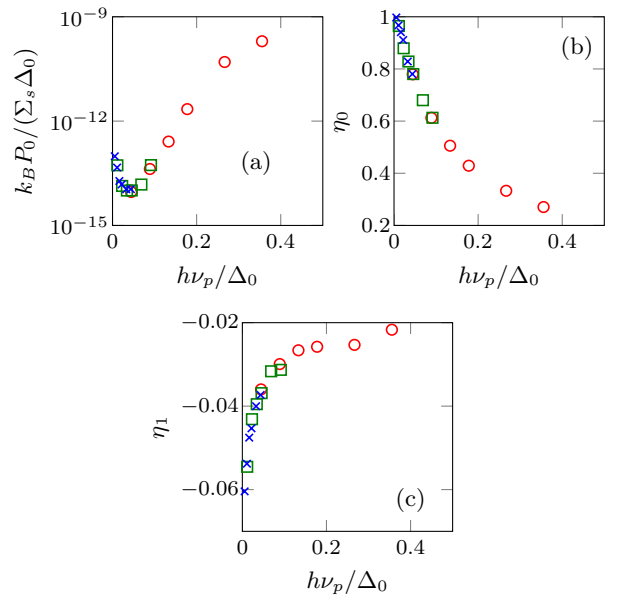
The dependence of the fit parameters $P_0$, $\eta_0$, and $\eta_1$ on probe photon energy is shown in \cref{fig:fitparamsvshnup}.
This includes calculations for
Nb (blue $\times$), Ta (green $\square$), and Al (red $\bigcirc$).
Scaling the energy of the photon by $\Delta_0$ demonstrates the monotonic behaviour for all the materials. $P_0$ shown in \cref{fig:P0vshnup} is
 scaled by the material-dependent $\Sigma_s$ to emphasize the commonality.

\begin{table}[tbh]
  \centering
  \ifhavepgf
  \begin{filecontents*}{tables/table3.dat}
Material    Delta0    Sigma P0
Al  180 3.22798e+10 0.00282829
Ta  700 1.17186e+14 13.0279
Nb  1470    3.82572e+15 3017.02
Mo  140 1.41609e+10 0.00295779
NbN 2560    1.29034e+16 31729.9
\end{filecontents*}
\pgfplotstabletypeset[%
    sort, sort key=Delta0, sort cmp=float <,
    column type=r,
    columns/Material/.style = {string type, column type=l},
    columns/Delta0/.style = {column name=$\Delta_0 \unitspace ( \si{\micro eV} )$, dec sep align},
    columns/Sigma/.style = {column name=$\Sigma_s \unitspace ( \si{W.m^{-3}.K^{-1}} )$, dec sep align},
    columns/P0/.style = {column name=$P_0 \unitspace ( \si{W.m^{-3}} )$, sci, sci zerofill, dec sep align}
  ]{tables/table3.dat}
  \else
  \begin {tabular}{lr<{\pgfplotstableresetcolortbloverhangright }@{}l<{\pgfplotstableresetcolortbloverhangleft }r<{\pgfplotstableresetcolortbloverhangright }@{}l<{\pgfplotstableresetcolortbloverhangleft }r<{\pgfplotstableresetcolortbloverhangright }@{}l<{\pgfplotstableresetcolortbloverhangleft }}%
\toprule Material&\multicolumn {2}{c}{$\Delta _0 \unitspace ( \si {\micro eV} )$}&\multicolumn {2}{c}{$\Sigma _s \unitspace ( \si {W.m^{-3}.K^{-1}} )$}&\multicolumn {2}{c}{$P_0 \unitspace ( \si {W.m^{-3}} )$}\\\midrule %
Mo&$140$&$$&$1$&$.42\cdot 10^{10}$&$2$&$.96\cdot 10^{-3}$\\%
Al&$180$&$$&$3$&$.23\cdot 10^{10}$&$2$&$.83\cdot 10^{-3}$\\%
Ta&$700$&$$&$1$&$.17\cdot 10^{14}$&$1$&$.30\cdot 10^{1}$\\%
Nb&$1\,470$&$$&$3$&$.83\cdot 10^{15}$&$3$&$.02\cdot 10^{3}$\\%
NbN&$2\,560$&$$&$1$&$.29\cdot 10^{16}$&$3$&$.17\cdot 10^{4}$\\\bottomrule %
\end {tabular}%

  \fi
  \caption{Superconductor cooling model material parameter $\Sigma_s$, and knee power
$P_0$ at $h\nu\tu{probe} = \SI{16}{\micro eV}$ and $\tau_l/\tau_0^\phi = 1$, obtained from fits to results of the full nonequilibrium calculation.}
  \label{tab:coolingparams}
\end{table}
\Cref{tab:coolingparams} summarizes parameters derived from the modelling to describe sub-gap photon interactions in all of the materials studied.
The table shows $\Sigma_s$ and knee parameter $P_0$ at $h\nu\tu{probe} = \SI{16}{\micro eV}$ and $\tau_l/\tau_0^\phi = 1$. The values of
$\Sigma_s$ shown can also be used for estimates of $T_N^*$ for a direct pair-breaking signal as we will discuss in the next section
with an appropriate energy-dependent $\eta$.

\subsection{Effects of direct pair-breaking}\label{sec:eta-pb}
Here we discuss results for a direct pair-breaking signal with $2\Delta < h\nu\tu{signal} < 10\Delta$.
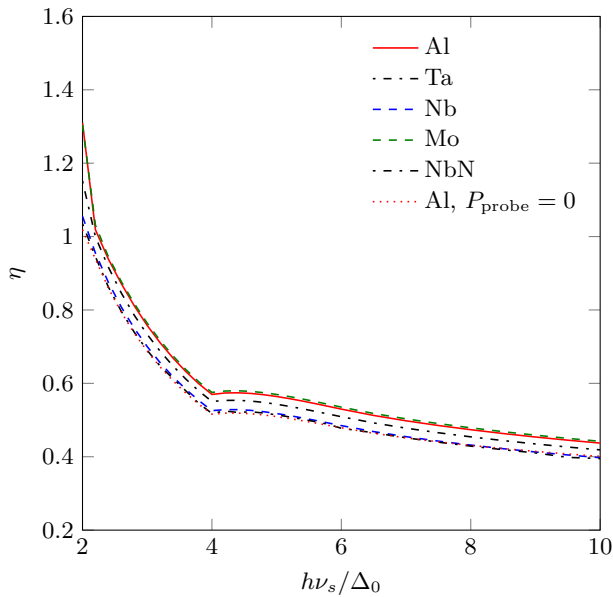
\begin{figure}[tbh]
  \centering
  \tikzsetnextfilename{paper_figure5}
%
%
\begin{tikzpicture}

\begin{axis}[%
width=\figurewidth,
height=\figureheight,
scale only axis,
xmin=2,
xmax=10,
xlabel={$h\nu_s / \Delta_0$},
ymin=0.2,
ymax=1.6,
ylabel={$\eta$},
legend style={fill=none,draw=none,legend cell align=left}
]
\addplot [color=red,solid]
  table[row sep=crcr]{2	1.31000764537545\\
2.2	1.01845809893807\\
2.4	0.938521163857953\\
2.6	0.871651114390457\\
2.8	0.811306614245544\\
3	0.757696536029332\\
3.2	0.710495794513104\\
3.4	0.668774440734805\\
3.6	0.631662540027403\\
3.8	0.598447069121334\\
4	0.569795689150285\\
4.2	0.573697987709557\\
4.4	0.574268904022625\\
4.6	0.572543010597626\\
4.8	0.569031944521559\\
5	0.564156052961345\\
5.2	0.55828359895227\\
5.4	0.551708747557652\\
5.6	0.544656048892089\\
5.8	0.537294799653617\\
6	0.529755439248702\\
6.5	0.512526721965661\\
7	0.497921584703369\\
7.5	0.485133911675829\\
8	0.473672300581623\\
8.5	0.46325513204089\\
9	0.4537647815019\\
9.5	0.44510578854052\\
10	0.43718443944339\\
};
\addlegendentry{Al};

\addplot [color=black,dash pattern=on 1pt off 3pt on 3pt off 3pt]
  table[row sep=crcr]{2	1.15168656152137\\
2.2	0.999278916665836\\
2.4	0.916541408170429\\
2.6	0.846076905865247\\
2.8	0.785655300239083\\
3	0.733285392768257\\
3.2	0.687459850531315\\
3.4	0.647024510588676\\
3.6	0.611081397919361\\
3.8	0.578921872986836\\
4	0.550583544810053\\
4.2	0.553486506905487\\
4.4	0.553765290407446\\
4.6	0.551768177193415\\
4.8	0.548039635801636\\
5	0.543040324596801\\
5.2	0.537132158682017\\
5.4	0.530591653748344\\
5.6	0.523627483343772\\
5.8	0.516396229454864\\
6	0.509015268520111\\
6.5	0.492147464821446\\
7	0.477869210267734\\
7.5	0.465382313800448\\
8	0.454206329175617\\
8.5	0.444062998505642\\
9	0.434832639216046\\
9.5	0.426418797751746\\
10	0.418727847937618\\
};
\addlegendentry{Ta};

\addplot [color=blue,dashed]
  table[row sep=crcr]{2	1.05523844235199\\
2.2	0.954698338206948\\
2.4	0.875144888250527\\
2.6	0.80782569468755\\
2.8	0.750124198838144\\
3	0.700116382869632\\
3.2	0.656359546731835\\
3.4	0.617750550120013\\
3.6	0.583431474420503\\
3.8	0.552725418134704\\
4	0.525622801669722\\
4.2	0.528279042625443\\
4.4	0.528411087098312\\
4.6	0.526371078237044\\
4.8	0.522683993249918\\
5	0.51779050704056\\
5.2	0.51203637177523\\
5.4	0.505685513741431\\
5.6	0.498936932773609\\
5.8	0.49193981378227\\
6	0.484805855520678\\
6.5	0.468497985320414\\
7	0.454679514749406\\
7.5	0.442586471017765\\
8	0.431759354227888\\
8.5	0.421931453965473\\
9	0.412987102719893\\
9.5	0.404832864631291\\
10	0.397378187268851\\
};
\addlegendentry{Nb};

\addplot [color=black!50!green,dashed]
  table[row sep=crcr]{2	1.30487482864406\\
2.2	1.0308487009957\\
2.4	0.945127004819414\\
2.6	0.877202027835379\\
2.8	0.817572878304528\\
3	0.764020477733462\\
3.2	0.716578289850072\\
3.4	0.674553213441103\\
3.6	0.637146584358069\\
3.8	0.603681158548086\\
4	0.575247529460932\\
4.2	0.579075125442111\\
4.4	0.579775992906227\\
4.6	0.578093213610328\\
4.8	0.574604937134737\\
5	0.569744474336716\\
5.2	0.563873255943491\\
5.4	0.557284531851018\\
5.6	0.550205438873097\\
5.8	0.542808104436905\\
6	0.535230117249047\\
6.5	0.517907248934684\\
7	0.503216090150144\\
7.5	0.490348787592804\\
8	0.478812012784677\\
8.5	0.468323425543038\\
9	0.45876564852716\\
9.5	0.450043326776825\\
10	0.442062637861295\\
};
\addlegendentry{Mo};

\addplot [color=black,dash pattern=on 1pt off 3pt on 3pt off 3pt]
  table[row sep=crcr]{2	1.03524125417529\\
2.2	0.944967986386826\\
2.4	0.862575275909936\\
2.6	0.796222130422514\\
2.8	0.739348853515082\\
3	0.690058893844286\\
3.2	0.646930244865753\\
3.4	0.612845447501509\\
3.6	0.579018997490729\\
3.8	0.544784162902795\\
4	0.518065464159352\\
4.2	0.524622965327817\\
4.4	0.520745920243221\\
4.6	0.522665654111676\\
4.8	0.518993793788214\\
5	0.514134141615868\\
5.2	0.508427765262163\\
5.4	0.502134938218285\\
5.6	0.49545185179726\\
5.8	0.488525543207474\\
6	0.477498170874762\\
6.5	0.465327274751422\\
7	0.451647104800429\\
7.5	0.439672194746404\\
8	0.428949431694541\\
8.5	0.41921567091252\\
9	0.410356517369021\\
9.5	0.39831240827013\\
10	0.394894969010874\\
};
\addlegendentry{NbN};

\addplot [color=red,dotted]
  table[row sep=crcr]{2	1.01898333414563\\
2.2	0.938444444524199\\
2.4	0.859902217232588\\
2.6	0.793627917016262\\
2.8	0.73686592848053\\
3	0.687687108546754\\
3.2	0.644666262859015\\
3.4	0.606708010266708\\
3.6	0.572970735666077\\
3.8	0.542787805394098\\
4	0.516140144175138\\
4.2	0.518765953179202\\
4.4	0.518996387237421\\
4.6	0.517174700302659\\
4.8	0.513807721278034\\
5	0.509316238899962\\
5.2	0.504016674134104\\
5.4	0.498159473649304\\
5.6	0.491929556527425\\
5.8	0.485465196405351\\
6	0.478873094084421\\
6.5	0.463901532042968\\
7	0.451402858588749\\
7.5	0.440600254126654\\
8	0.431006186967812\\
8.5	0.422369579535834\\
9	0.41455834913121\\
9.5	0.407480509600658\\
10	0.401050833743665\\
};
\addlegendentry{Al, $P\tu{probe} = 0$};

\end{axis}
\end{tikzpicture}%
  \caption{Quasiparticle generation efficiency $\eta\tu{power}$ against
signal photon energy $h\nu\tu{signal}$ for different materials.
Calculated for $T_b = 0.1\,T_c$, $P\tu{probe} = \SI{2E3}{W.m^{-3}}$,
$h\nu\tu{probe} = \SI{16}{\micro eV}$, $P\tu{signal} = \SI{2}{W.m^{-3}}$, and $\tau_l/\tau_0^\phi = 1$.}
  \label{fig:etavsmaterial}
\end{figure}
\Cref{fig:etavsmaterial} shows how $\eta$ varies
 with signal photon energy.
$\eta$ decreases from unity in the range $h\nu\tu{signal} = 2\Delta$ to $4\Delta$,
as phonons emitted by quasiparticles scattering towards the gap escape the thin film.
Above $h\nu\tu{signal} = 4\Delta$, the phonons emitted in scattering can break additional Cooper pairs, increasing the quasiparticle generation efficiency.
The magnitude of the increase depends on the phonon trapping factor~\cite{Guruswamy2014}.
Measurements of the response of a Ta KID~\cite{Neto2014} at signal frequencies close to the gap energy show these characteristic features.
The figure shows that when $h\nu\tu{signal}$, $T_b$ and $\tau_l$ are scaled by the relevant material parameters,
the quasiparticle generation efficiency is the same for all materials.
\Citet{Kozorezov2000} predicted that the quasiparticle generation efficiency could be categorized into three classes of superconductors.
The superconductors included in this work, which fall within either the second
(Ta, Nb, NbN) or third (Al, Mo) class, have a material-independent $\eta \approx 0.6$.
This conclusion differs somewhat from that of \citeauthor{Zehnder1995}~\cite{Zehnder1995}.
We understand this as due to the current work examining the steady-state response to constant incoming power, and scaling all
relevant parameters by material characteristics, whereas \citeauthor{Zehnder1995}'s work examined
the time-dependent response including quasiparticle out-diffusion from a localized hot-spot, and chose
a fixed cutoff time of \SI{10}{ns} to calculate the quasiparticle generation efficiency unscaled by material.
We emphasize that for the low energy photons considered in the present work localized hot-spot formation (and consequent
gap suppression)
does not occur for the signal powers studied.

The other effect shown in \cref{fig:etavsmaterial} is the enhancement of $\eta$ by the readout power
due to an increase in the average energy of generated quasiparticles. For
 $h\nu\tu{signal}\simeq 2\Delta$, it can be seen that $\eta>1$ which may seem unreasonable.
The enhancement is due to multiple photon absorption by the quasiparticles i.e. both signal and probe, and
the effect is most-pronounced near $h\nu\tu{signal} = 2\Delta$.
This results in a higher average excess quasiparticle energy when the
readout power is present alongside the signal compared to when only the signal is present,
and also produces the small variation in $\eta$ between materials.
The absorbed readout power $P\tu{probe}$ and readout frequency $h\nu\tu{probe}$ are not scaled by material parameters in this figure, unlike the signal.
When calculated with $P\tu{probe} = 0$, the signal quasiparticle generation efficiency is identical for all materials and is unity at $h\nu\tu{signal} = 2\Delta$.

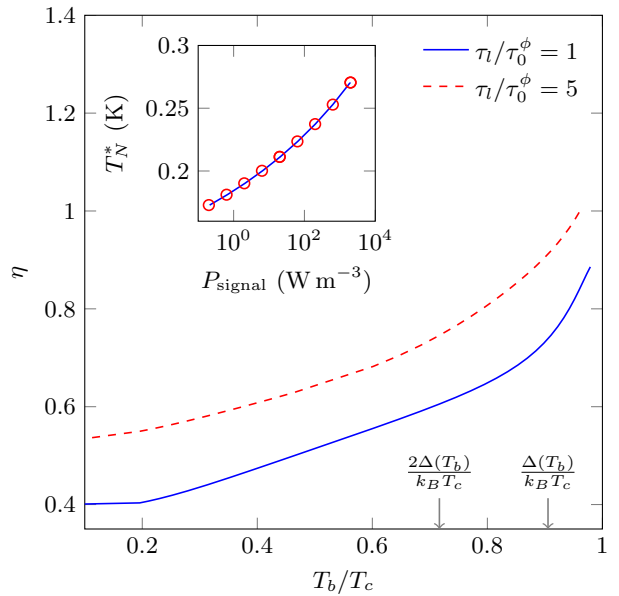
\begin{figure}[tbh]
  \centering

  \ifhavepgf
  \tikzset{
    every pin/.append style = {font=\footnotesize},
  }
  \tikzexternaldisable
  \savebox\mainplot{
%
%
\begin{tikzpicture}

\begin{axis}[%
width=\figurewidth,
height=\figureheight,
scale only axis,
xmin=0.1,
xmax=1,
xlabel={$T_b / T_c$},
ymin=0.35,
ymax=1.4,
ylabel={$\eta$},
legend style={fill=none,draw=none,legend cell align=left},
extra description/.code = {\node[coordinate,pin=above:{$\frac{\Delta(T_b)}{k_B T_c}$}] at (axis cs:0.90541,0.35) {}; \node[coordinate,pin=above:{$\frac{2\Delta(T_b)}{k_B T_c}$}] at (axis cs:0.716584,0.35) {};}
]
\addplot [color=blue,solid]
  table[row sep=crcr]{0.978975717387908	0.885755866550897\\
0.973246555731377	0.872596414458202\\
0.966776393968909	0.856605457825604\\
0.959540210389521	0.838711230837663\\
0.951508722383635	0.819816514728873\\
0.942647726148289	0.800642812374515\\
0.932917264421616	0.781779063916112\\
0.922270569262108	0.763515013249424\\
0.910652705694886	0.746087849953977\\
0.897998810572464	0.729585123190671\\
0.88423177317491	0.713955602665333\\
0.8692591295465	0.699163651227848\\
0.852968823063344	0.685069688778739\\
0.835223285803441	0.671531233644484\\
0.815850955069389	0.658371442058235\\
0.794633728756082	0.64541241482982\\
0.771287709899638	0.632433535249648\\
0.745432275174536	0.619188715291358\\
0.716537492264072	0.605394743960458\\
0.683828003652881	0.590632028410747\\
0.646089503210883	0.57434851121987\\
0.601222222387457	0.555616246333566\\
0.465596196277566	0.500486229217489\\
0.452611495075723	0.495245022906515\\
0.438321784131586	0.489486224921354\\
0.422340011343383	0.483064111303257\\
0.404050340182314	0.475731836713687\\
0.401450912285926	0.474680094920034\\
0.382367399847579	0.467077024851896\\
0.355038208165533	0.456307382888794\\
0.35116360566094	0.454798126233166\\
0.315643257455737	0.44121738015194\\
0.311166663384449	0.439557640429849\\
0.278844630335644	0.427907739688971\\
0.252312653532293	0.41905247573619\\
0.230212618735397	0.412327864584171\\
0.211557870511573	0.407217221503339\\
0.195622668598848	0.403360809989007\\
0.1	0.401050833743665\\
};
\addlegendentry{$\tau_l/\tau_0^\phi = 1$};

\addplot [color=red,dashed]
  table[row sep=crcr]{0.959540210389521	0.997149635839301\\
0.951508722383635	0.981381536912501\\
0.942647726148289	0.965564111589459\\
0.932917264421616	0.949911963847027\\
0.922270569262108	0.934230669476933\\
0.910652705694886	0.9185976406061\\
0.897998810572464	0.902937574762823\\
0.88423177317491	0.887144103020601\\
0.8692591295465	0.871120048650735\\
0.852968823063344	0.854763527717406\\
0.835223285803441	0.837975360658216\\
0.815850955069389	0.820689211966392\\
0.794633728756082	0.802845797749659\\
0.771287709899638	0.784395400492369\\
0.745432275174536	0.765331794562694\\
0.716537492264072	0.74565444332892\\
0.683828003652881	0.725317674271291\\
0.646089503210883	0.704262991784678\\
0.601222222387457	0.682189700736506\\
0.465596196277566	0.629619046721061\\
0.452611495075723	0.625212757702279\\
0.438321784131586	0.620442244220538\\
0.422340011343383	0.615202139863186\\
0.404050340182314	0.6092845857078\\
0.401450912285926	0.608438527075579\\
0.382367399847579	0.602365431251906\\
0.355038208165533	0.593810580255668\\
0.35116360566094	0.592611310533902\\
0.315643257455737	0.581792855724905\\
0.311166663384449	0.580455607269157\\
0.278844630335644	0.571022676090189\\
0.252312653532293	0.56366215266275\\
0.230212618735397	0.557881344702915\\
0.211557870511573	0.553305161998236\\
0.195622668598848	0.54968761779646\\
0.1	0.534640682633122\\
};
\addlegendentry{$\tau_l/\tau_0^\phi = 5$};

\end{axis}
\end{tikzpicture}%
}

  \savebox\insetplot{%
    \setlength{\figureheight}{0.13\textwidth}%
    \setlength{\figurewidth}{0.13\textwidth}%
%
%
\begin{tikzpicture}

\begin{axis}[%
width=\figurewidth,
height=\figureheight,
scale only axis,
xmode=log,
xmin=0.1,
xmax=10000,
xminorticks=true,
xlabel={$P\tu{signal}\unitspace (\si{W.m^{-3}})$},
ymin=0.16,
ymax=0.3,
ylabel={$T_N^*\unitspace (\si{K})$}
]
\addplot [color=red,only marks,mark=o,mark options={solid},forget plot]
  table[row sep=crcr]{0.2	0.172861056620116\\
0.632455532033676	0.181120020001162\\
2	0.190186451584624\\
6.32455532033676	0.200181792598358\\
20	0.211251288760991\\
20	0.2112514652108\\
63.2455532033676	0.223572057567179\\
200	0.237359305673054\\
632.455532033676	0.252880531164869\\
2000	0.270469973210276\\
2000	0.270470391925672\\
};
\addplot [color=blue,solid,forget plot]
  table[row sep=crcr]{0.210778523998843	0.172861056620116\\
0.664265385253255	0.181120020001162\\
2.09171784634967	0.190186451584624\\
6.57911016039555	0.200181792598358\\
20.6614083601046	0.211251288760991\\
20.6614090170486	0.2112514652108\\
64.7586621890685	0.223572057567179\\
202.495382770189	0.237359305673054\\
631.515427030509	0.252880531164869\\
1963.93756825825	0.270469973210276\\
1963.93845750462	0.270470391925672\\
};
\end{axis}
\end{tikzpicture}%
%
  }
  \tikzexternalenable
  \fi

  \tikzsetnextfilename{paper_figure6}
  \begin{tikzpicture}
  \node[anchor=south west, inner sep=0] (myplot) at (0,0) {\usebox\mainplot};
  \begin{scope}[x={(myplot.south east)}, y={(myplot.north west)}]
    \node[anchor=north west, inner sep=0] at (0.15,0.95) {\usebox\insetplot};
  \end{scope}
  \end{tikzpicture}

  \caption{The quasiparticle generation efficiency $\eta$
 as a function of reduced temperature $T_b/T_c$.
The calculation is for for Al, with $P\tu{probe} = 0$,
$P\tu{signal} = \SI{2}{W.m^{-3}}$, $h\nu\tu{signal} = 10\,\Delta(T_b)$, and two values of the phonon escape time. The inset shows $T^*_N$
calculated using the full nonequilibrium model (red $\bigcirc$) and using \cref{eq:p_model} (blue line) as a function of $P\tu{signal}$ with $T_b = 0.1\,T_c$ and $\tau_l/\tau_0^\phi=1$.
}
  \label{fig:etavst}
\end{figure}

\Cref{fig:etavst} shows the variation in $\eta$ as the bath temperature is changed.
As $T_b$ increases, the scattering between the more numerous thermal quasiparticles and phonons determines
the structure of the driven distribution, rather than scattering between the excess quasiparticles and phonons.
This means the power flow from phonons to bath does not have the same peaked structure and instead more closely
resembles the difference between two thermal distributions, as earlier shown in \cref{fig:convergence}.
This type of excess quasiparticle distribution means the average energy of generated quasiparticles
increases without increasing the recombination and pair breaking rates.
The figure also indicates temperatures for which $ k_BT_b = 2\Delta(T_b)$ and $=\Delta(T_b)$, above which
we expect thermal and nonequilibrium distributions to interact strongly.
As $T_b/T_c\to 1$ we see that $\eta\to 1$ in agreement with the two-temperature model valid in that regime. We find that $T_N^*$ is well-accounted for by \cref{eq:p_model},
with the value of $\eta$ shown, for $T_b/T_c \le 0.7$. The inset shows $T^*_N$
calculated using \cref{eq:p_model} for $\tau_l/\tau_0^\phi=1$ as a function of $P\tu{signal}$ with $T_b = 0.1\,T_c$.

\begin{figure}[tbh]
  \centering
  \tikzsetnextfilename{paper_figure7}
%
%
\begin{tikzpicture}

\begin{axis}[%
width=\figurewidth,
height=\figureheight,
scale only axis,
xmin=0,
xmax=15,
xlabel={$\tau_l / \tau_0^\phi$},
ymin=0.3,
ymax=0.65,
ylabel={$\eta$},
legend style={at={(0.97,0.03)},anchor=south east,fill=none,draw=none,legend cell align=left}
]
\addplot [color=red,solid]
  table[row sep=crcr]{0.5	0.337194974680318\\
0.75	0.373607209430507\\
1	0.401050833743665\\
1.5	0.439940324387469\\
2	0.466327198435135\\
3	0.500002543436582\\
4	0.520657820825922\\
5	0.534640682633122\\
6	0.544742485879527\\
7	0.552382381120321\\
8	0.558364510740717\\
9	0.563174292423818\\
10	0.567126624854461\\
11	0.57043064277864\\
12	0.573234562776204\\
13	0.575642690188654\\
14	0.577733988070637\\
15	0.579566018369582\\
};
\addlegendentry{Al};

\addplot [color=black,dash pattern=on 1pt off 3pt on 3pt off 3pt]
  table[row sep=crcr]{0.5	0.335808905889374\\
0.75	0.372373820206358\\
1	0.399899187929053\\
1.5	0.438884535275286\\
2	0.465326372689867\\
3	0.499104789477051\\
4	0.519842557770978\\
5	0.533894762700862\\
6	0.544053432424213\\
7	0.551741945752143\\
8	0.55776626910739\\
9	0.562613308626043\\
10	0.566598960700223\\
11	0.569933182168258\\
12	0.572764757887997\\
13	0.57519847601933\\
14	0.577313641731816\\
15	0.579168117306323\\
};
\addlegendentry{Ta};

\addplot [color=blue,dashed]
  table[row sep=crcr]{0.75	0.363594489559929\\
1.5	0.434528380633894\\
2	0.462478592622421\\
3	0.497710191515749\\
4	0.519062271158229\\
5	0.53341579523556\\
6	0.54373766947147\\
7	0.551523356773843\\
8	0.557606200545147\\
9	0.562495467726116\\
10	0.56650765541396\\
11	0.56986121131056\\
12	0.572706880970102\\
13	0.575152152173384\\
15	0.579135959177578\\
};
\addlegendentry{Nb};

\addplot [color=black!50!green,dashed]
  table[row sep=crcr]{0.5	0.33768013907002\\
0.75	0.374143753276151\\
1	0.401616067549934\\
1.5	0.440513079597889\\
2	0.466892047402858\\
3	0.500533694219856\\
4	0.521149105393907\\
5	0.535096720799853\\
6	0.545165034254054\\
7	0.552777512526897\\
8	0.558733487689017\\
9	0.563521438817811\\
10	0.567452497340784\\
11	0.570738367437248\\
12	0.573524255648068\\
13	0.575916704682908\\
14	0.577992185516083\\
15	0.579810238922753\\
};
\addlegendentry{Mo};

\addplot [color=black,dash pattern=on 1pt off 3pt on 3pt off 3pt]
  table[row sep=crcr]{4	0.517822573418876\\
6	0.543101787352865\\
7	0.551038365440434\\
8	0.557233096479753\\
9	0.56219937549148\\
11	0.569671658406041\\
12	0.572551919738198\\
13	0.575019429681745\\
14	0.577162738453743\\
15	0.579041892108483\\
};
\addlegendentry{NbN};

\end{axis}
\end{tikzpicture}%
  \caption{Quasiparticle generation efficiency $\eta$ for materials studied as a function of phonon escape time $\tau_l/\tau_0^\phi$.
Calculation uses $T_b = 0.1\,T_c$, $P\tu{probe} = 0$, $P\tu{signal} = \SI{2}{W.m^{-3}}$, and $h\nu\tu{signal} = 10\,\Delta$.}
  \label{fig:etavstaul}
\end{figure}
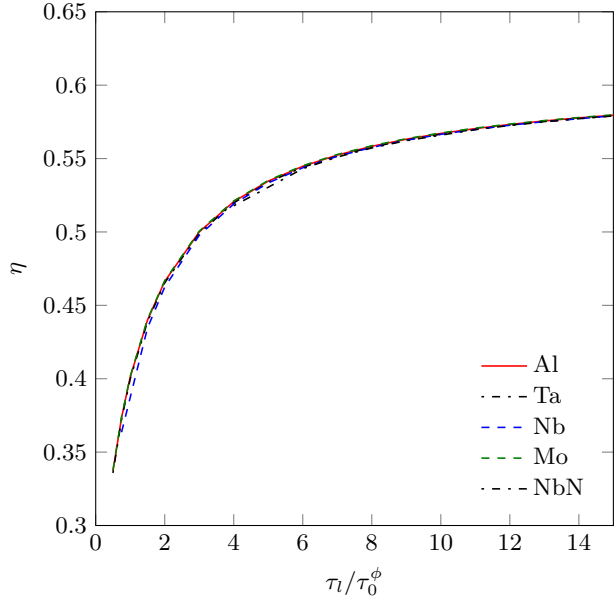

\Cref{fig:etavstaul} shows the dependence of $\eta$ on the phonon trapping factor $\tau_l/\tau_{pb}$ for $T_b = 0.1\,T_c$.
As the phonon escape time $\tau_l/\tau_0^\phi$ increases, the quasiparticle generation efficiency $\eta \rightarrow 0.6$
 for all materials examined.
This conclusion is in agreement with earlier Monte Carlo calculations~\cite{Kurakado1982,Hijmering2009}
for infinite superconductors or where $\Omega \ge 2\Delta$ phonon loss was ignored.

\section{Conclusions}\label{sec:conclusions}
We have described solutions of the coupled kinetic equations that
calculate steady-state nonequilibrium quasiparticle and phonon distributions in a number of
technologically important superconducting thin-films
driven by sub-gap and pair-breaking photons. In particular we consider low energy signal photon interactions $h\nu\tu{signal}\le 10\Delta$,
where localized heating and gap-suppression can be ignored.
We have calculated numerical parameters for these
 superconductors that can be used in a simple analytical expression to
describe the power-flow between quasiparticles and phonons. This expression allows
straight-forward estimates of the effective quasiparticle temperatures for both sub-gap and pair-breaking drives to
approximate the nonequilibrium behaviour without resorting to a full numeric solution of the coupled kinetic equations.
 The analytical expression is shown to give
a good account of full solutions of the detailed equations for a wide range of powers and
bath temperatures.
This is relevant for predicting the behaviour of thin-film nonequilibrium superconducting detectors.
We defined a low-energy quasiparticle generation efficiency for constant absorbed pair-breaking power and calculated
detailed numerical solutions as a function of photon energies and bath temperature.
 We have shown that key parameters determining quasiparticle generation efficiency are the signal frequency and the phonon trapping factor.
The enhancement of signal quasiparticle generation efficiency by absorbed readout power is demonstrated and explained as multiple-photon absorption (signal and probe) by the quasiparticles and Cooper pairs.
We estimate this effect can result in an increase $\eta$ of up to 20\% at signal photon energies of $h\nu\tu{signal} = 2\Delta$.
The most sensitive detectors currently in development may also be able to distinguish between phonon trapping factors using results presented here.
For the low-energy photon interactions studied
we also show that the quasiparticle generation efficiency is material independent.

\bibliography{paper_noneq}   %
\end{document}